# Magnetic ordering above room temperature in the sigma-phase of $Fe_{66}V_{34}$

Jakub Cieslak[1], Benilde F. O. Costa[2], Stanislaw M. Dubiel[1*], Michael Reissner[3] and Walter Steiner[3]

[1]Faculty of Physics and Compute Science, AGH University of Science and Technology, 30-059 Krakow, Poland, [2] CEMDRX Department of Physics, University of Coimbra, 3000-516 Coimbra, Portugal, [3]Institute of Solid State Physics, Vienna University of Technology, 1040 Wien, Austria

PACS: 75.30.Cr, 75.50.Bb, 76.80.+y, 77.80.Bh

**Abstract**

Magnetic properties of four sigma-phase $Fe_{100-x}V_x$ samples with $34.4 \leq x \leq 55.1$ were investigated by Mössbauer spectroscopy and magnetic measurements in the temperature interval 5 – 300 K. Four magnetic quantities viz. hyperfine field, Curie temperature, magnetic moment and susceptibility were determined. The sample containing 34.4 at% V was revealed to exhibit the largest values found up to now for the sigma-phase for average hyperfine field, $<B>$ = 12.1 T, average magnetic moment per Fe atom, $<\mu>$ = 0.89 $\mu_B$, and Curie temperature, $T_C$ = 315.5 K. The quantities were shown to be strongly correlated with each other. In particular, $T_C$ is linearly correlated with $<\mu>$ with a slope of 406.5 K/$\mu_B$, as well as $<B>$ is so correlated with $<\mu>$ yielding 14.3 T/$\mu_B$ for the hyperfine coupling constant.

---

- Corresponding author: dubiel @novell.ftj.agh.edu.pl

## 1. Introduction

Among over 50 examples of the σ-phases (tetragonal unit cell - structure type $D_{4h}^{14}$ $P4_2/mnm$) known to exist in binary alloy systems, only two viz. σ-FeCr and σ-FeV exhibit magnetic order [1]. Magnetism of the σ-phase in the Fe-Cr system can be termed as weak itinerant, with the highest Curie temperature $T_C$ = 39 K, and the largest average magnetic moment per Fe atom $<\mu>$ = 0.29 $\mu_B$ found in the sample with 45 at% Cr [2]. Exchange interaction in the σ-phase of the Fe-V system is much stronger. Here, the highest value of $T_C$ = 267.5 K was recorded for a sample with 37.3 at%V [3], and the largest value of $<\mu>$ = 0.92 $\mu_B$ was reported for a sample containing 36 at% V [4]. However, the latter was not directly measured, but estimated from the average hyperfine field using a conversion factor (hyperfine coupling constant) of 14.7 T/$\mu_B$. The largest value of 0.62 $\mu_B$ was derived from magnetization measurements for a σ-$Fe_{63.8}V_{34.2}$ sample [5].

It should be noted that the difference in the values of the above discussed magnetic quantities i.e. $T_C$ and $<\mu>$ found for the σ-phase in the Fe-Cr and Fe-V systems, does not only originate from the difference in composition at which these extreme values were measured. In fact, if one makes such comparison of the magnetic quantities for similar compositions, one finds a significant difference for the two systems.. In particular, the difference in the magnetic moment found for the two systems increases linearly with the increase of Fe content in favour of FeV i.e. the rate of increase is higher for FeV than the one for FeCr [6,7].

The present investigation was motivated by our previous results recorded on a series of σ-$Fe_{100-x}Cr_x$ samples with 45 ≤ x ≤ 50. The results were obtained using magnetization measurements and Mössbauer-effect techniques [2,8].

First, we found that the values of $<\mu>$ determined from magnetization measured in external magnetic fields, $B_a$, up to 15 T were shifted by ~0.1 $\mu_B$ to higher values in comparison with their counterparts derived from similar curves recorded in an external field of 1 T [6,7]. This difference follows from the ill-defined extrapolation condition of the magnetization measured in the external magnetic field of 1 T to $B_a = 0$ T.

Second, we have revealed that the correlation of the average hyperfine field with $<\mu>$ is not linear. In other words, the conversion factor (hyperfine coupling constant) changes continuously its value with composition between ~9 T/$\mu_B$ for x close to 45, and ~18 T/$\mu_B$ for x close to 50 [8].

In the light of the above given information, it is of interest to verify whether or not similar effects can be found for the σ-FeV phase, which is in fact more appropriate for testing the above-mentioned characteristics because the composition range of the σ-phase occurrence in Fe-V is by a factor ~6 larger than the one in the Fe-Cr system [9].

## 2. Sample preparation

Master alloys of α-$Fe_{100-x}V_x$ with a nominal composition x = 36, 40, 48 and 60 were prepared by melting appropriate amounts of Fe (99.95% purity) and V (99.5% purity) in an arc furnace under protective argon atmosphere. The ingots received after melting were next solution treated at 1273 K for 72 hours followed by a water quench. The chemical composition was determined on the homogenized samples by electron probe microanalysis as x = 34.4, 39.9, 47.9 and 55.1. The transformation into the σ-phase was performed by annealing the ingots at $T_a = 973$ K for 25 days. The verification of the α to σ phase transformation was done by recording room temperature X-ray and neutron diffraction patterns [10]. It should be noted that the σ-phase samples contained some small fraction of V-rich precipitates, most likely in form of carbides. These precipitates were not detected in the neutron diffraction patterns. However, it is known that carbide nucleation often precedes σ-phase nucleation [11].

### 3. Results and discussion

#### 3.1. Curie temperature

#### 3.1.1. Mössbauer-effect measurements

$^{57}$Fe Mössbauer spectra were recorded in the temperature interval of 4.2 – 300 K with a standard spectrometer and a $^{57}$Co/Rh source of the γ-rays. The temperature of a sample, which was placed in a cryostat, was kept constant to within ±0.2 K. The shape of the spectra sensitively depends on a sample composition and, for a given composition, on temperature. To illustrate the former, a set of the spectra recorded at 4.2 K for x = 34.4, 39.9, 47.9 and 55.0 are presented in Fig. 1. The influence of temperature on the shape of the spectrum is shown in Fig. 2 for x = 34.4.

The spectra were fitted with the hyperfine field distribution (HFD) method to get the distribution of the hyperfine field, *P(B)*. It was assumed that the hyperfine field was linearly correlated with the isomer shift and with the quadrupole splitting. Some examples of the *P(B)*-curves obtained in this way are presented in Fig. 3. By their integration, the average hyperfine field, *<B>*, was calculated. From the temperature dependence of *<B>*, the Curie temperature, $T_{C1}$, was estimated for each sample. In particular, from the *<B>(T)* plot obtained for the σ-Fe$_{65.6}$V$_{34.4}$ sample, which is shown in Fig. 4, $T_C$ = 324 K was derived (The non-zero value of *<B>* at $T \geq 324$ K follows from the fact that the spectrum in the paramagnetic phase is not a single-line, while it was fitted with the HFD method, so the departure from the single-line was accounted for by a small hyperfine field ). To our best knowledge, this is the highest value of the Curie temperature ever recorded for a σ-FeV sample.

#### 3.1.2 Magnetization versus temperature measurements

Measurements of magnetization, *M*, were performed with a vibrating sample magnetometer (VSM) in a constant magnetic field, $B_a$, versus temperature, *T*. Typical curves are shown in the upper part of Fig. 5. By plotting *dM/dT* versus *T* – see lower part of Fig. 5, the Curie points were derived. In particular, the value of $T_{C2}$ = 306.6 K was found for the σ-$Fe_{65.6}V_{34.4}$ sample. It is smaller than the corresponding value evaluated from the average hyperfine field, but it is known that such difference between the two methods may occur as far as the Curie point is concerned. In any case, the record high-value of $T_{C1}$ found from the *<B>(T)* plot has been confirmed with the *M(T)* plot. $T_C$-values being the arithmetic average over $T_{C1}$ and $T_{C2}$ are presented in Fig. 6, together with other values obtained in this study, as well as those found in the literature [3,5-7,12-15]. A non-linear dependence on the compound composition can be readily seen.

**3.2 Magnetic moment and suceptibility**

Magnetic moment was derived from the magnetization measurements in an external magnetic field $B_a$ ≤ 9 T at constant temperature. Typical *M($B_a$)* – curves are presented in Fig. 7. By extrapolation of the linear part of the *M($B_a$)* – curve recorded at 5 K to $B_a$ = 0 T, the saturation magnetization value $M_s$ was found, and from it the average magnetic moment per Fe atom, *<μ>*, was evaluated. Its non-linear dependence on the compound composition can be seen in Fig. 8, where both the values found in this study as well as those known from the literature are plotted. An upwards shift of the presently found data relative to those from the literature can be readily seen. The reason for the shift, also revealed previously for σ-FeCr samples [2], is a result of the extrapolation condition of the magnetization curves to $B_a$= 0 T. Our extrapolation procedure was from much larger $B_a$- values, hence more reliable. This, in the light of a non-saturating character of the *M($B_a$)*- curves. had resulted in different zero-field saturation values, hence *<μ>*. The *M($B_a$)* data recorded for x = 39.9 and 47.9 at *T > $T_{C2}$* were used to determine the effective paramagnetic moment, $\mu_{eff}$, which was next used to verify the Rhodes-Wohlfarth criterion for itinerant magnetism. For this purpose the susceptibility, $\chi$ = $M/B_a$ , was determined. An example of the data is displayed in Fig. 9. The linear part of the inverse susceptibility was fitted to the Curie –Weiss formula

$$\chi = \chi_o + \frac{C}{T - \Theta} \qquad (1)$$

The best-fit values of the parameters obtained in such a way are displayed in Table 1. From $C$ the $\mu_{eff}$ – values were derived. The Curie temperature, $T_C$, and the average magnetic moment per Fe atom, $<\mu>$, and the average hyperfine field, $<B>$, are included in this table, too. From the data, and in particular, from the $\mu_{eff}/\mu_s$ ratio, $\mu_s$ being the saturation moment, it is clear that according to the Rhodes-Wohlfarth criterion, the magnetism of the studied system has itinerant character.

### 3.3. Curie point – magnetic moment correlation

It is well known that magnetizations as a function of the valence electron number per atom of 3d transition metal systems form the so-called Slater-Pauling curve. Similarly, the Curie temperatures of these alloys also exhibit a Slater-Pauling like behaviour [16].

Though the overall shape of the latter is like the former, the two differ in details. It follows from this fact that the Curie temperature depend not only on the magnetization but also on the magnetic exchange interactions coefficients that are characteristic of a given alloy system.

Indeed, recent theoretical calculations carried out with the KKR-CPA-LDA method for several 3d transition binary alloys have clearly demonstrated that the Curie temperature – magnetic moment relationship is characteristic of a given system [17]. In other words, the two quantities are, in general, not linearly correlated.

As shown in Fig. 10 for the σ-FeV system, $T_C$ is very well linearly correlated with the magnetic moment, $<\mu>$. The slope is equal to 406.5 K/$\mu_B$, and is significantly different from the one found recently for the σ-phase in the Fe-Cr system (i.e. 205 K/$\mu_B$) [2,8]. A linear correlation between the two magnetic quantities was also observed in the α-phase of the Fe-V system of similar composition as the one in the presently studied samples [4]. The slope for the latter is, however, equal to 741 K/$\mu_B$ i.e. almost twice the one in the σ-FeV system. For comparison, the $T_C/\mu$ value for a pure iron is equal to 470 K/$\mu_B$.

This comparison clearly shows that the $T_C$ - $<\mu>$ relationship depends not only on the alloy system, but for a given system it depends on its crystallographic structure.

### 3.4. Average hyperfine field - magnetic moment correlation

As illustrated in Fig. 11 for the σ-FeV compounds, the average hyperfine field, $<B>$ - is linearly correlated with the average magnetic moment per Fe atom, $<\mu>$, with the slope of 14.3 T/$\mu_B$. Such correlation is rather strange and unexpected for this system as according to theoretical calculations performed on bcc-Fe-systems [18-20], only a part of the hyperfine field viz. the one due to a polarization of the core electrons is proportional to the Fe-site magnetic moment, while the second part viz. the one due to the polarization of the conduction electrons, $B_{CEP}$, is not. In other words, a non-linear $<B>$ - $<\mu>$ relationship might be taken as evidence that a substantial contribution to the hyperfine field originates form the conduction electrons i.e. the system is magnetically itinerant. In the presently studied case, the Rhodes-Wohlfarth criterion speaks in favour of the latter. The linear $<B>$ - $<\mu>$ relationship means that also $B_{CEP}$ term is proportional to $<\mu>$. This conclusion must be, however, verified by theoretical calculations devoted to the investigated system.

The presently found correlation between $<B>$ and - $<\mu>$ is also of a practical meaning as it can be used as the reference plot permitting a unique transformation between the two quantities. It is linear and different than the one found previously for the σ-Fe-Cr system [8]. The latter feature gives clear evidence that the $<B>$ - $<\mu>$ relationship is not universal, but it is rather characteristic of a given structure and system.

## 4. Conclusions

The results obtained in this study permit drawing the following conclusions:

(a) σ-$Fe_{65.6}V_{34.4}$ sample shows the strongest magnetic properties ever found for the sigma-phase as measured in terms of the Curie temperature, magnetic moment and magnetic hyperfine field.

(b) Curie temperature and magnetic moment per Fe atom show nonlinear decrease with V content.

(c) Curie temperature and the average hyperfine field are linearly correlated with the magnetic moment. The latter is rather unexpected for an itinerant system and prompts theoretical calculations.

**Acknowledgement**

Dr. Jan Żukrowski is thanked for melting the samples.

**References**

[1] E. O. Hall and S. H. Algie, Metall. Rev., **11** (1966) 61

[2] J. Cieslak, M. Reissner, W. Steiner and S. M. Dubiel, J. Magn. Magn. Mater., **272-276** (2004) 534; Phys. Stat. Sol (a), (2008)/DOI 10.1002/pssa.200723618

[3] H. H. Ettwig and W. Pepperhoff, Arch. Eisenhuttenwes., **43** (1972) 271

[4] A. M. van der Kraan, D. B. de Mooij and K. H. J. Buschow, Phys. Stat. Sol. (a), **88** (1985) 231

[5] Y. Sumimoto, T. Moriya, H. Ino and F. E. Fujita, J. Phys. Soc. Jpn., **35** (1973) 461

[6] D. A. Read and E. H. Thomas, IEEE Trans. Magn., MAG-**2** (1966) 415

[7] D. A. Read, E. H. Thomas and J. B. Forsythe, J. Phys. Chem. Solids, **29** (1968) 1569

[8] J. Cieślak, B. F. O. Costa, S. M. Dubiel, M. Reissner and W. Steiner, J. Phys.,: Condens. Matter., **17** (2005) 2985

[9] O. Kubaschewski, Iron-Binary Phase Diagrams, Springer Verlag, 1982, Berlin

[10] J. Cieślak, M. Reissner, S. M. Dubiel, J. Wernisch and W. Steiner, J. Alloys Comp., **20** (2008) 20

[11] R. Blower and G. J. Cox, J. Iron Steel Inst., August 1970, 769

[12] D. Parsons, Nature, 185 (1960) 839

[13] M. V. Nevitt and P. A. Beck, Trans. AIME, May 1955, 669

[14] M. Mori and T. Mitsui, J. Phys. Soc. Jpn., 22 (1967) 931

[15] M. V. Nevitt and A. T. Aldred, J. Appl. Phys., 34 (1963) 463

[16] H. P. Wijn, in *Magnetic Properties of Metals*, 1991, ed. R. Poerschke, Berlin, Springer

[17] C. Takahashi, M. Ogura and H. Akai, J. Phys.: Condens. Matter, **19** (2007) 365233

[18] R. E. Watson and A. J. Freeman, Phys. Rev., **123** (1961) 2027

[19] M. E. Elzain, D. E. Ellis and D. Guenzburger, Phys. Rev. B, **34** (1986) 1430

[20] B. Lingren and J. Sjöström, J. Phys. F: Metal. Phys., **18** (1988) 1563

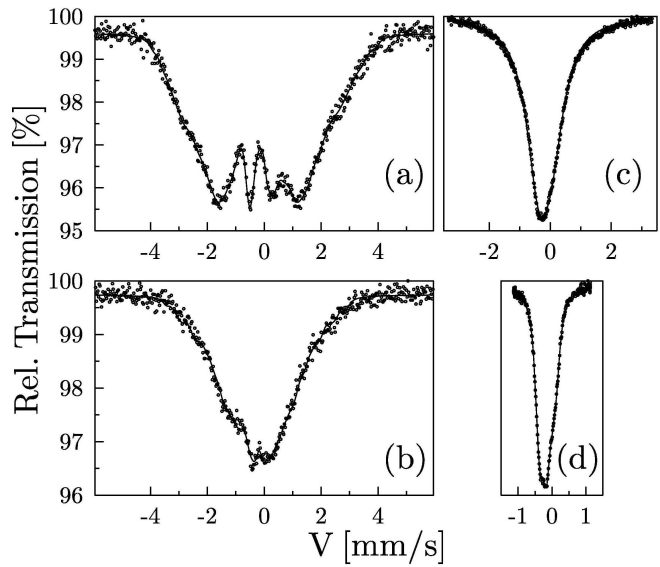

Fig. 1
$^{57}$Fe Mössbauer spectra recorded at 4.2 K on the σ-Fe$_{100-x}$V$_x$ samples: (a) x = 34.4, (b) x = 39.9 (c) x = 47.9 and (d) x = 55.1

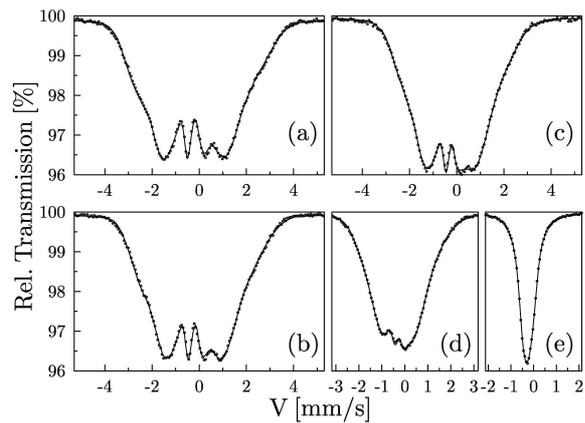

Fig. 2
$^{57}$Fe Mössbauer spectra recorded on the σ-Fe$_{65.6}$V$_{34.4}$ sample at: (a) 4.2 K, (b) 120 K, (c) 195 K, (d) 269 K and (e) 330 K.

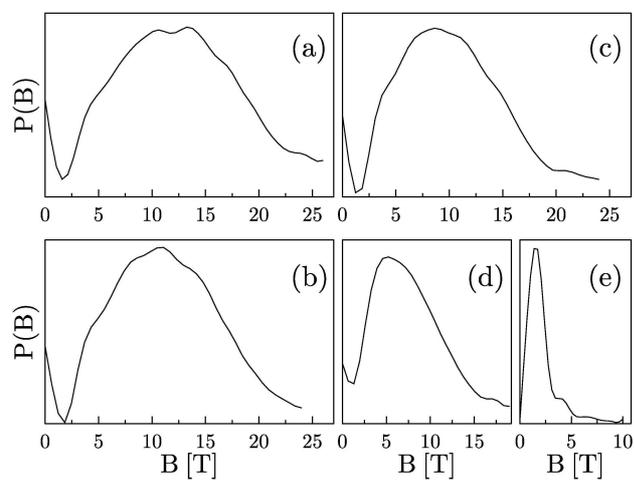

Fig. 3
Distribution of the hyperfine field for the σ-Fe$_{65.6}$V$_{34.4}$ sample obtained from the spectra recorded at: (a) 41 K, (b) 120 K, (c) 195 K, (d) 269 K and (e) 330 K.

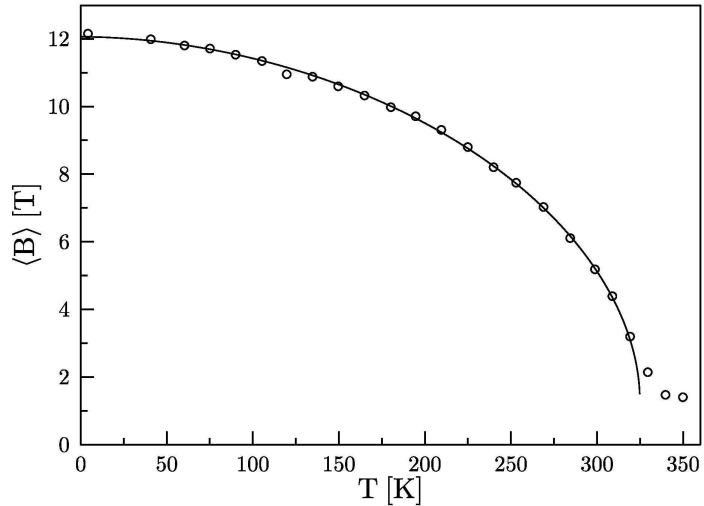

Fig. 4
Average hyperfine field, <B>, versus temperature, T, for the σ-Fe$_{65.6}$V$_{34.4}$ sample with the estimated value of the Curie point T$_c$ = 325 K.

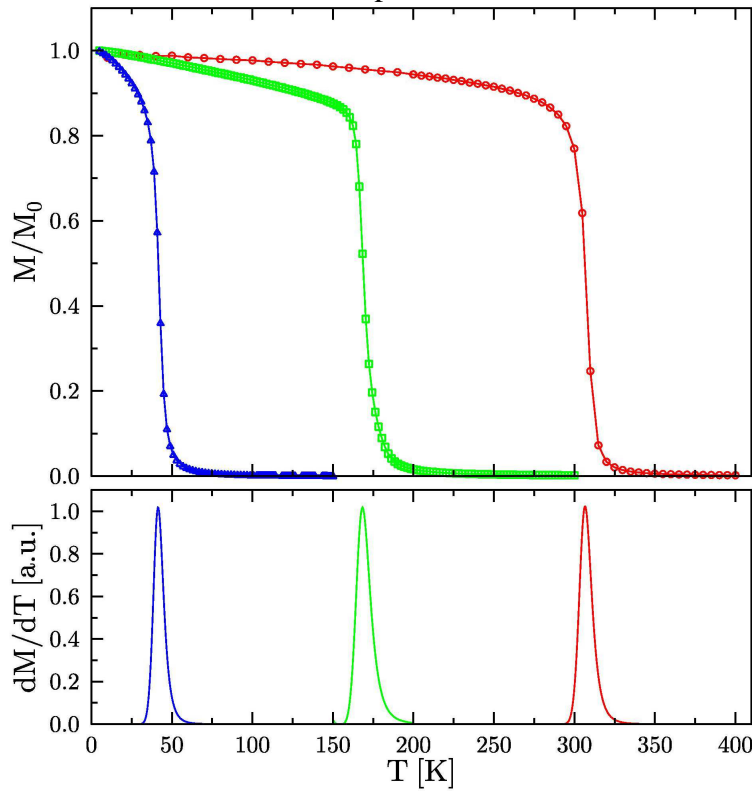

Fig. 5
Normalized magnetization curves, M/M$_o$ (M$_o$ being the value of M at 4.2 K and B$_a$ = 100, 700 and 176 Oe, respectively) versus temperature T (top) and dM/dT curves (bottom) for three of the studied samples (x = 34.4, 39.9.0 and 47.9 in sequence from right to left).

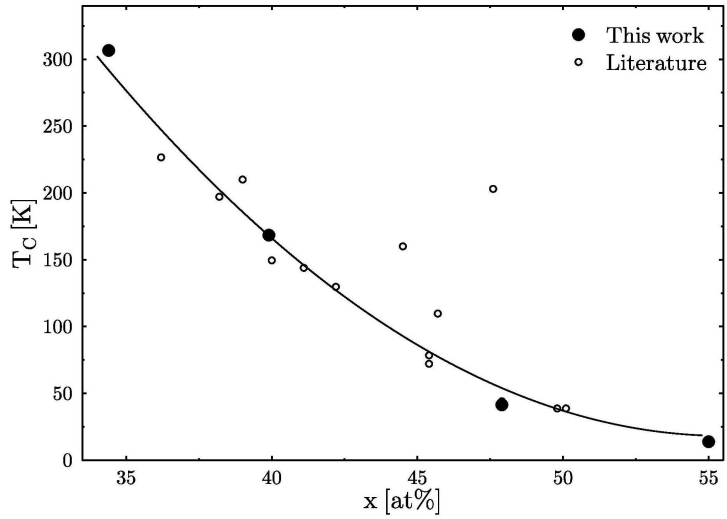

Fig. 6
Curie temperature, $T_C$ versus V content, x for the σ-FeV compounds. The solid line is a guide to the eye. Empty symbols are taken from literature [3,5-7,12-15].

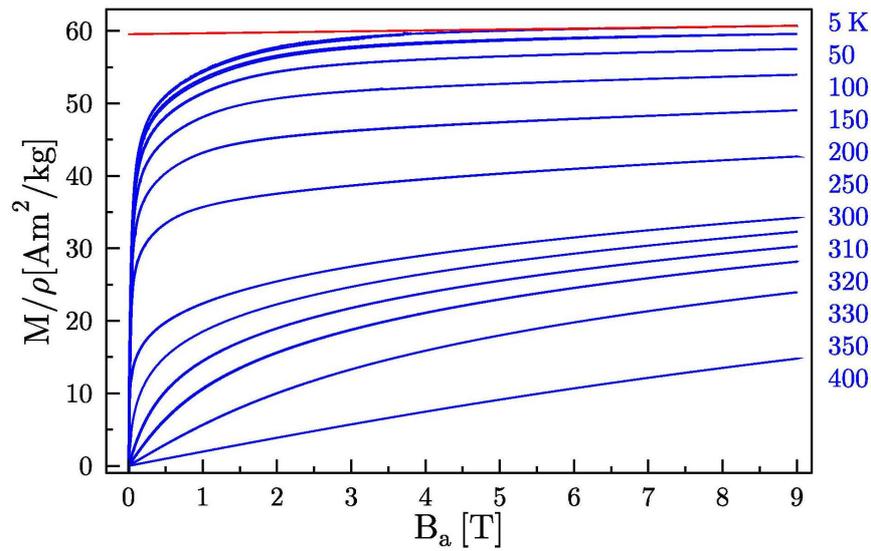

Fig. 7
Magnetization curves recorded on the σ-$Fe_{65.6}V_{34.4}$ sample versus external magnetic field $B_a$ for various temperatures shown. Saturation magnetization was determined by extrapolation of the linear part of the curve measured at 5 K to $B_a = 0$ T as indicated.

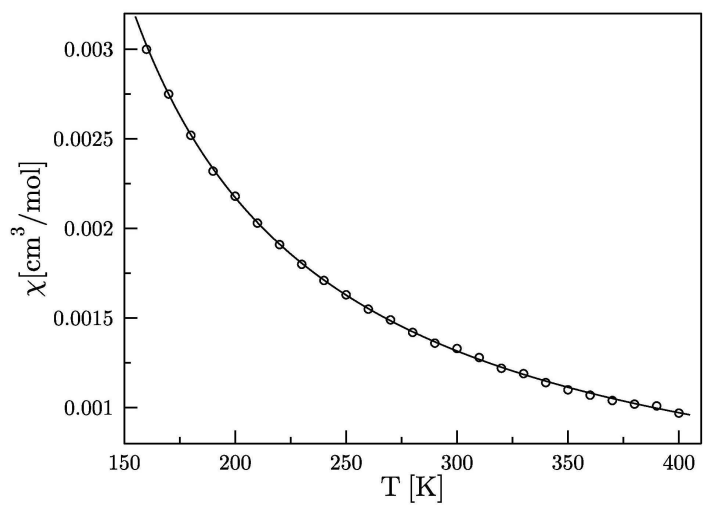

Fig. 8 Susceptibility versus temperature for the σ-Fe$_{66.6}$V$_{34.4}$ sample. The solid line is the best-fit to the data in terms of equation (*).

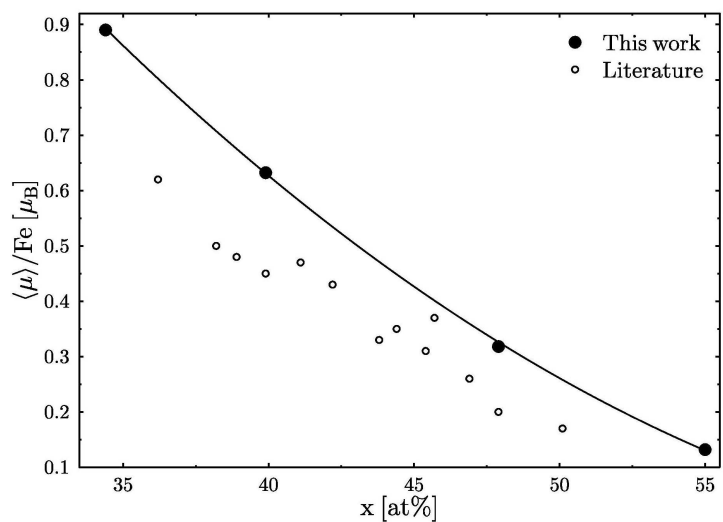

Fig. 9
Magnetic moment per Fe atom, <μ>, for the σ-FeV compounds versus V content, x. The solid line is to guide the eye. Empty symbols represent data taken from literature [3,5-7,12-15].

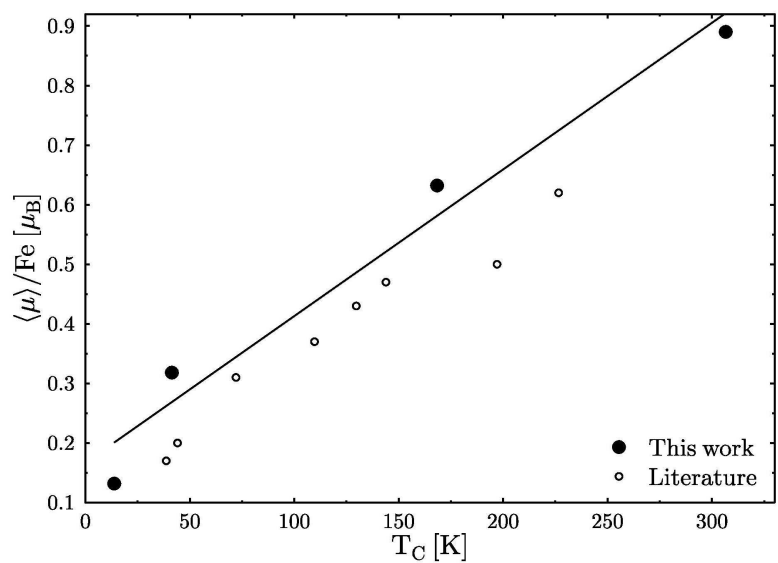

Fig. 10
Linear correlation between the average magnetic moment per Fe atom, <μ>, and the Curie temperature, $T_C$ for the σ-FeV compounds. The solid line is the best-fit to the data. The extreme points span over the whole concentration range of the σ-phase existence.

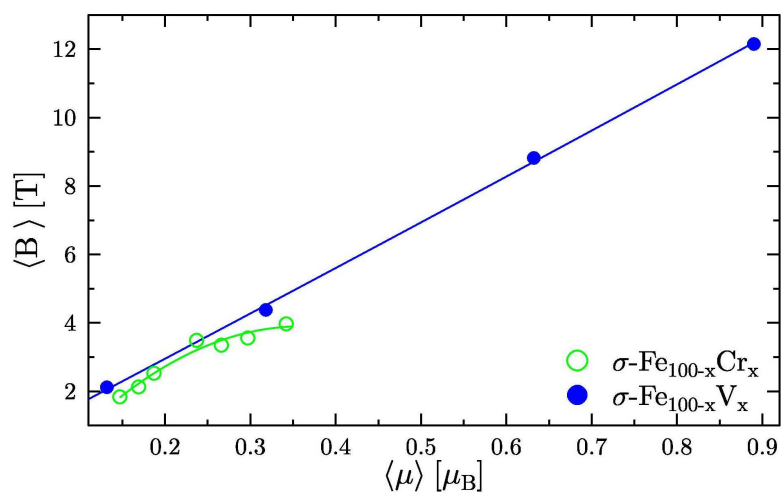

Fig. 11
Average hyperfine field, <B>, versus average magnetic moment per Fe atom, <μ>, in the σ-phase in the Fe-Cr [8] and Fe-V systems. The solid lines are the guide to the eye.

**Table 1** Magnetic quantities determined for the studied σ-Fe$_{100-x}$V$_x$ samples. The Curie temperature as determined from the Mössbauer measurements is denoted by $T_{C1}$, that from the magnetization measurements by $T_{C2}$, the average hyperfine field by $<B>$, the average magnetic moment per Fe atom by $<\mu>$, the magnetic saturation moment, $\mu_s$, the effective paramagnetic moment by $\mu_{eff}$ and the paramagnetic Curie temperature by Θ.

| x [at%] | T$_{C1}$ [K] | T$_{C2}$ [K] | $<B>$ [T] | $<\mu>$ [$\mu_B$] | $\mu_{eff}$ [$\mu_B$] | $\mu_s$ [$\mu_B$] | Θ [K] |
|---|---|---|---|---|---|---|---|
| 34.4 | 324.0 | 306.6 | 12.2 | 0.89 | - | 0.58 | - |
| 39.9 | 149.0 | 168.4 | 8.8 | 0.63 | 2.162 | 0.38 | 237 |
| 47.9 | 40.0 | 41.4 | 4.4 | 0.32 | 2.005 | 0.165 | 68 |
| 55.0 | 17.0 | 13.8 | 2.1 | 0.13 | - | 0.06 | - |